\documentclass[runningheads]{llncs}
\usepackage[T1]{fontenc}
\usepackage{graphicx}
\usepackage{url}
\usepackage{hyperref}
\usepackage{amsfonts}
\usepackage{amsmath}

\begin{document}
\title{Spatiotemporal Distillation via Recurrent Bottlenecks for Aortic Tracking}

\author{Dexter Wen Jie Teo\inst{1,2}\thanks{Code available at \url{https://github.com/dexterteo4/aortic-temporal-distillation}.} \and
Nairouz Shehata\inst{2} \and
Herve Lombaert\inst{2}}

\authorrunning{D. Teo et al.}
\institute{College of Computing and Data Science, Nanyang Technological University, Singapore \and
Department of Computer and Software Engineering, Polytechnique Montreal, Canada
\\ Corresponding author: \email{WTEO030@e.ntu.edu.sg}}
\maketitle
\begin{abstract}
Cardiac cine-MRI serves as a direct visual indicator of cardiovascular hemodynamics by capturing the continuous wall motion of the aorta. Quantifying these dynamic structural changes across the cardiac cycle is essential for measuring aortic distensibility, a primary marker of arterial stiffness. However, standard 2D segmentation networks focus on each frame independently. Consequently, when rapid systolic flow temporarily obscures the aorta's boundaries, this lack of continuous context results in frame-to-frame tracking dropouts and boundary inconsistencies. Spatiotemporal ($2\text{D}+t$) networks can enforce temporal consistency across the sequence but suffer from a scarcity of expert annotations. To address this, we present a semi-supervised spatiotemporal ($2\text{D}$ to $2\text{D}+t$) knowledge distillation framework exploiting the cardiac cycle. The framework distills a spatial teacher's expertise into a spatiotemporal student network by executing a dynamic latent interception, pairing a recurrent spatiotemporal bottleneck with a residual spatial bypass. Our model selection strategy applies a baseline validation threshold ($\text{DSC} \ge 0.50$) prior to selecting the epoch that maximizes anatomical consistency. This strategy enables the spatiotemporal student model to achieve superior surface tracking accuracy ($\text{NSD@1mm} = 92.3\% \pm 0.2\%$) and high structural reliability ($\text{Frac}_{2\text{CC}} = 99.2\% \pm 0.6\%$), reducing population-wide structural anomalies by over 56\% compared to a 2D nnU-Net baseline.

\keywords{Cine-MRI \and Knowledge distillation \and Semi-supervised learning \and Spatiotemporal segmentation \and Aortic tracking}
\end{abstract}
\section{Introduction}
Tracking the ascending and descending aorta across the full cardiac cycle is essential for evaluating cardiovascular function. In short-axis views, the aortic arch intersects the imaging plane as two distinct, circular components. Continuous cross-sectional area measurements yield biophysical metrics like aortic distensibility and pulse wave velocity~\cite{cecelja2022aortic}. These markers act as primary indicators of arterial stiffness, cardiovascular disease, and aneurysm risk~\cite{townsend2015recommendations,wentland2014review}. However, manual annotation of these cine-MRI sequences across dozens of frames per cardiac cycle is highly labor-intensive and suffers from observer variability~\cite{jaffre2023deep}. This variability compromises the reproducibility of the tracked segmentation boundaries across consecutive frames~\cite{bratt2021deep}.

Standard 2D Convolutional Neural Networks (CNNs) such as U-Net~\cite{ronneberger2015u} achieve high volumetric overlap~\cite{chen2020deep} but evaluate slices independently. They mostly ignore temporal dynamics and fail to preserve topological consistency across time~\cite{mosinska2018beyond}. This results in anatomically inaccurate segmentations containing internal holes or disconnected fragments~\cite{painchaud2020cardiac}. This tracking inconsistency is heavily driven by rapid systolic blood acceleration, which causes fading visibility of the vessel boundaries~\cite{storey2004flow}. Lacking temporal context, static 2D networks frequently fail to detect the obscured ascending aorta entirely or mistakenly segment adjacent, brighter vascular structures in its place~\cite{painchaud2020cardiac,ruijsink2020fully}. Furthermore, frames at the start and end of sequences suffer from low-contrast flow fade-out, leading to geometric under-segmentation along ambiguous vessel borders~\cite{chen2020deep}.

Prior recurrent models propagate shape context from sparse anchor frames~\cite{bai2018recurrent}, but require manual labels for every sequence and remain susceptible to topological drift. While semi-supervised Knowledge Distillation (KD) scales efficiently using unannotated videos~\cite{tarvainen2017mean,radosavovic2018data}, standard 2D teachers propagate artifact-induced hallucinations into the pseudo-label pool, destabilizing student training.

To overcome these limitations, we introduce a parameter-efficient, anatomically prioritized spatiotemporal distillation framework for label-efficient 2D aortic tracking. Pseudo-labels generated by a static 2D Teacher network are passed through an offline, deterministic topological sanitization operator to remove biologically incorrect fragments. These clean labels are distilled into a spatiotemporal student network. We hypothesize that pairing a recurrent spatiotemporal bottleneck with a temporal smoothness loss allows the student network to leverage sequential context. This sequential context enables the model to look across adjacent frames, smoothing over transient dropouts to preserve boundary continuity throughout the cardiac cycle.

The primary contributions of this paper are as follows:
\begin{itemize}

\item \textbf{Decoupled Training Strategy:} A two-stage student-teacher framework that separates spatial learning from temporal tracking, using an offline structural mask correction to enforce anatomical constraints on unannotated pseudo-labels.

\item \textbf{Latent Spatiotemporal Bottleneck:} A parameter-efficient recurrent block integrated into the student's deepest hidden layer to capture sequence context without expanding model parameters.

\item \textbf{Temporal Smoothness Regularization:} A differentiable temporal loss function that penalizes rapid tracking flicker and boundary jitter between consecutive frames while allowing smooth, natural physiological motion.
\end{itemize}

\section{Methodology}
Our spatiotemporal distillation method leverages temporal context from neighboring frames to enforce sequence-wide anatomical consistency on unannotated data. Our approach operates in two stages: a fully supervised spatial teacher training phase coupled with a deterministic topological correction operator, followed by a sequence-wide student distillation phase that optimizes frame-to-frame dependencies via a temporal smoothness objective.

\subsection{Problem Formulation and Proposed Approach}
A cine-MRI sequence $X \in \mathbb{R}^{T \times 1 \times H \times W}$ maps to binary masks $Y \in \{0,1\}^{T \times 1 \times H \times W}$ across the cyclic cardiac timeline. To guarantee anatomical validity, each frame must contain exactly two connected components ($N=2$) representing the ascending and descending aorta.

To handle data scarcity, we leverage three distinct pools: a small end-diastolic labeled dataset ($\mathcal{D}_L$), a large unannotated distillation pool ($\mathcal{D}_D$), and an isolated baseline test set ($\mathcal{D}_T$). Our decoupled framework operates in two phases: Phase I trains a spatial teacher on $\mathcal{D}_L$ to generate pseudo-labels across $\mathcal{D}_D$, which are structurally filtered to enforce $N=2$. Phase II distills these anatomically stabilized labels into a spatiotemporal student network, forcing it to exploit sequence context to maintain tracking continuity.

\subsection{Phase I: Teacher Optimization and Geometric Label Correction}
The 2D spatial teacher network is optimized on $\mathcal{D}_L$ using a joint spatial objective balancing regional volumetric overlap with edge-focused class difficulties:
\begin{equation}
    \mathcal{L}_{\text{teacher}} = \mathcal{L}_{\text{Dice}}(Y, \hat{Y}) + \lambda \mathcal{L}_{\text{Focal}}(Y, \hat{Y})
\end{equation}
where $\hat{Y}$ represents predicted soft probability maps, $Y$ is the expert reference mask, and $\lambda$ is a balancing hyperparameter. To prioritize structural validity, checkpoint selection uses a validation topological gatekeeper. Candidate checkpoints must clear a validation floor ($\text{DSC} \ge \tau_{\text{dice}}$). The framework then maximizes the anatomical feasibility rate ($\text{Frac}_{2\text{CC}}$) across the validation split, using DSC only as a tiebreaker to reject epochs with fragmented segmentations.

Once optimized, the teacher runs inference across $\mathcal{D}_D$ to generate predictions, binarized via threshold $\tau_{\text{bin}}$ to construct raw pseudo-labels $M_t = \mathbb{I}(\hat{Y}_t > \tau_{\text{bin}})$. To clean these masks before student distillation, we execute an offline, deterministic \textbf{three-pass correction strategy} under an 8-connectivity rule. 

In \textbf{Pass 1 \& 2 (Verification and Pruning)}, frames possessing exactly two connected components are verified as correct. If a frame displays fragmentation ($N > 2$), the mask is automatically pruned to retain only its two largest valid anatomical bodies, and its frame index is appended to a valid anchor pool $\mathcal{G}$. 

In \textbf{Pass 3 (Bidirectional Circular Stitching)}, for frames with severe dropouts ($N < 2$), the framework locates the nearest stable reference $t^* \in \mathcal{G}$ using a circular distance function:
\begin{equation}
    t^* = \operatorname*{argmin}_{g \in \mathcal{G}} \left\{ \min \left( |g - t|, \, T - |g - t| \right) \right\}
\end{equation}
If frame $t$ retains one component ($N=1$), the centroid of this surviving blob is used to identify which aortic component is missing relative to $M_{t^*}$, and that missing component blob is copied in full into $M_t$. For completely empty frames ($N=0$), the entire mask $M_{t^*}$ is copied into $M_t$. Because $t^*$ is selected as the temporally closest valid frame, physical spatial displacement is minimal; furthermore, Phase II temporal smoothness regularization ($\mathcal{L}_{\text{smooth}}$) prevents the student from overfitting to local shape discontinuities.

\subsection{Phase II: Spatiotemporal Student Architecture and Objectives}
During Phase II, the student model leverages a hardware-efficient \textbf{batch-folding} mechanism to retain sequence context without 3D convolutions. Given an input sequence tensor $X \in \mathbb{R}^{B \times T \times 1 \times H \times W}$ from $\mathcal{D}_D$ (where $B$ is batch size), batch and temporal dimensions are collapsed via a flat view transformation: $X_{\text{folded}} \in \mathbb{R}^{(B \times T) \times 1 \times H \times W}$. This representation is fed directly into a 2D spatial attention U-Net \cite{oktay2018attention} backbone warm-started with the pre-trained weights from the Phase I teacher. To preserve structural capacity parity, the student replicates the teacher's deep encoder-decoder profile.

\begin{figure}[t]
    \centering
    \includegraphics[width=1\textwidth]{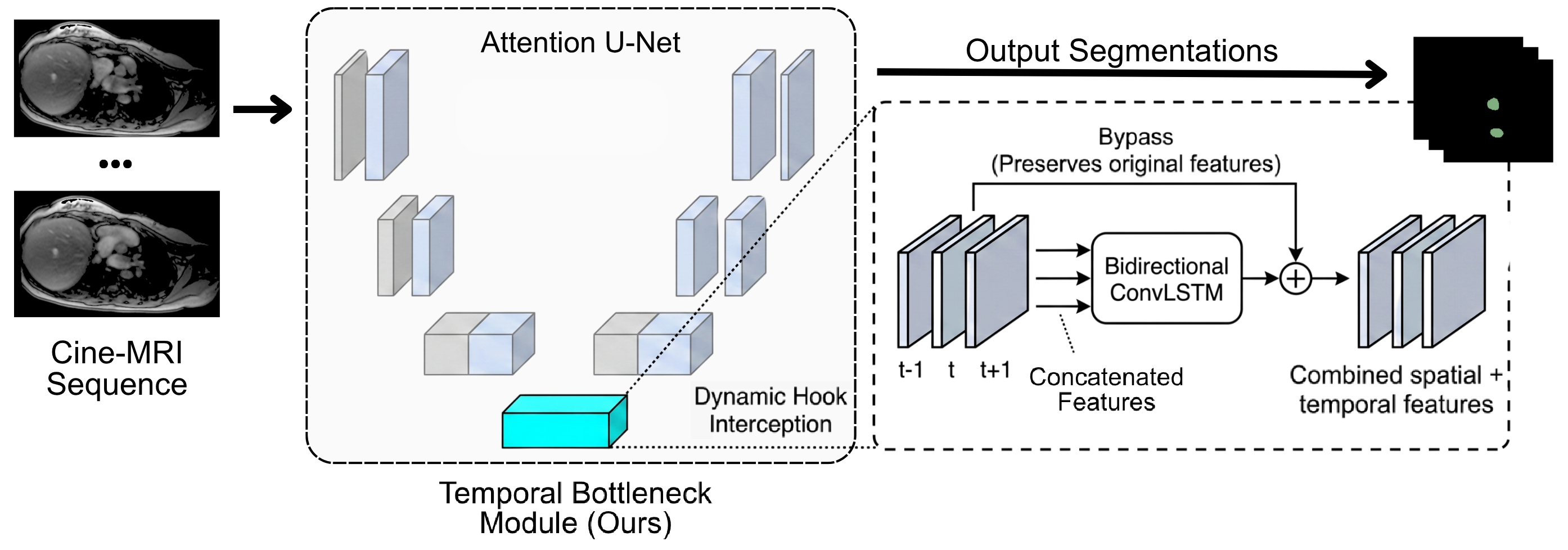}
    \caption{Overview of the spatiotemporal student architecture. A dynamic hook routes deep features through a bidirectional ConvLSTM to capture sequence context. This temporal stream merges with a residual spatial bypass ($\oplus$) to preserve tracking continuity and enforce anatomical plausibility ($N=2$)  across the cardiac cycle.}
    \label{fig:student_architecture}
\end{figure}

To inject temporal coherence, we execute a \textbf{dynamic latent bottleneck interception} using forward hooks, as illustrated in Fig.~\ref{fig:student_architecture}. The framework intercepts activations at the deepest hidden layer immediately preceding the upsampling decoder blocks. This hidden tensor is then unfolded back into its true sequential spatiotemporal configuration: $H_{\text{bottleneck}} \in \mathbb{R}^{B \times T \times C_{\text{in}} \times H_b \times W_b}$. This feature space is projected down to a compact hidden channel space ($C_{\text{temp}}$) via a $1 \times 1$ convolution, yielding a compressed tensor $H_{\text{comp}}$. A parameter-efficient bidirectional Convolutional LSTM (BiConvLSTM) processes this compressed stream. It executes a forward pass from $t=0$ to $T-1$ and a backward pass from $t=T-1$ to $0$ to establish continuous temporal states and bridge tracking dropouts.

Forward and backward temporal states are concatenated, projected back to the native bottleneck scale, and standardized channel-wise. This regularized correction is injected into the original spatial feature map via a residual branch. To maintain the integrity of the pre-trained spatial backbone, this temporal branch is zero-initialized and scaled by a fixed coefficient $\alpha$:
\begin{equation}
H_{\text{enhanced}} = H_{\text{bottleneck}} + \alpha \cdot \text{tanh}\left( \text{Conv}_{1 \times 1}\left( \text{BiConvLSTM}(H_{\text{comp}}) \right) \right)
\end{equation}
These enhanced features are re-folded and passed through standard upsampling pathways, yielding final tracking probabilities $\hat{Y}_{\text{student}} \in \mathbb{R}^{B \times T \times 1 \times H \times W}$.

To optimize the student network on the sanitized pseudo-labels, we utilize a compound objective function balancing local spatial correctness with sequence-wide temporal stabilization:
\begin{equation}
\label{eq:student_loss}
\mathcal{L}_{\text{student}} = \mathcal{L}_{\text{spatial}} + \beta \mathcal{L}_{\text{smooth}}
\end{equation}
where $\beta$ represents the temporal smoothness regularization weight. Here, $\mathcal{L}_{\text{spatial}}$ is implemented as a frame-weighted Dice and Binary Cross-Entropy loss aggregated across the sequence clip. Within this spatial component, the baseline Dice and cross-entropy elements are weighted at a 1.0 to 0.5 ratio, respectively. Non-central frames are assigned a uniform weight ($w_t = 1$). In contrast, the temporal anchor frame at the clip center is assigned an amplified weight ($w_{\text{center}} = 5$) to ground localized temporal propagation context. 

To penalize tracking flicker and edge stutter without constraining valid physiological motion, we implement a differentiable temporal smoothness loss operating over soft class probabilities:
\begin{equation}
    \mathcal{L}_{\text{smooth}} = \frac{1}{T-2} \sum_{t=2}^{T-1} \left\| \hat{Y}_{t+1} - 2\hat{Y}_{t} + \hat{Y}_{t-1} \right\|_1
\end{equation}
The loss computes the mean $L_1$ norm of the discrete second derivative. It evaluates to zero during uniform velocity but penalizes sudden tracking anomalies. This penalty directly enforces structural continuity across the cardiac timeline.

\section{Results}

\subsection{Dataset and Temporal Curation}
We evaluate our framework on 437 cardiac cine-MRI scans acquired on a 1.5T Siemens Magnetom Aera scanner (\texttt{tfi3D-fs} sequence; 2.0~mm pixel spacing, 1.6~mm slice thickness, $Z=1$). To avoid demographic bias and data leakage, we partitioned the cohort at the patient level into a 412-patient distillation pool ($\mathcal{D}_D$) and an unseen 25-patient test set ($\mathcal{D}_T$) using stratified random sampling across age, body surface area, and biological sex. Model development and 5-fold cross-validation were conducted strictly within $\mathcal{D}_D$; $\mathcal{D}_T$ remained untouched and was used exclusively for final evaluation across all models.

Ground-truth annotations differ between training and testing:
\begin{itemize}
    \item \textbf{Teacher Training Set ($\mathcal{D}_L$):} Medical experts annotated ground-truth contours at end-diastole (ED) for 198 patients within $\mathcal{D}_D$. Because the ED frame index varies dynamically with heart rate, we manually inspected each video sequence to pinpoint the exact ED frame.
    \item \textbf{Full-Sequence Test Set ($\mathcal{D}_T$):} To evaluate tracking across all 948 continuous frames in the test set, candidate masks were initially generated using a 2D nnU-Net baseline, then manually edited and audited frame-by-frame to correct boundary errors and topological failures.
\end{itemize}

\subsection{Experimental Setup and Implementation Details}
Our framework is implemented in PyTorch~\cite{paszke2019pytorch} via MONAI~\cite{cardoso2022monai} using 5-fold patient-level cross-validation. Inputs are downsampled to $320 \times 320$ pixels. Sequences are drawn using a sliding temporal window ($T=11$ frames, stride = 1) with circular wrap-around at boundaries. 

The student network is optimized via AdamW for 70 epochs (patience = 14) using $1\times10^{-5}$ weight decay and a Cosine Annealing schedule decaying to $1\times10^{-6}$. We employ differential learning rates: $\eta_{\text{enc}} = 1\times10^{-5}$ to fine-tune the spatial backbone and $\eta_{\text{temp}} = 1\times10^{-4}$ for the new temporal bottleneck components. All batch normalization layers are frozen in evaluation mode throughout.

\paragraph{Hyperparameter Specification and Quality Criteria.}
Model hyperparameters were configured as follows (summarized in Table~\ref{tab:hyperparameters}): pseudo-label binarization threshold $\tau_{\text{bin}} = 0.50$, teacher focal loss weight $\lambda_{\text{focal}} = 0.50$, temporal residual scaling factor $\alpha = 0.10$, and compressed temporal bottleneck width $C_{\text{temp}} = 128$. For population-wide audits (Table~\ref{tab:population_results}), frame confidence was evaluated via a composite metric:
\begin{equation}
S_{\text{conf}} = 0.7\,c_{\text{fg}} + 0.3\,c_{\text{certainty}}
\end{equation}
where $c_{\text{fg}}$ denotes the mean predicted foreground probability and $c_{\text{certainty}} = 2 \cdot \operatorname{mean}(|p - 0.5|)$ measures class boundary certainty. A frame was flagged as \textit{low-confidence} when $S_{\text{conf}} < 0.85$. Frames exhibiting a connected-component count $N \neq 2$ were separately cataloged as structural anomalies.

\begin{table}[h]
\centering
\caption{Summary of framework hyperparameters and operational thresholds.}
\label{tab:hyperparameters}
\resizebox{0.8\textwidth}{!}{%
\begin{tabular}{lll}
\hline
\textbf{Parameter} & \textbf{Symbol / Variable} & \textbf{Configured Value} \\ \hline
Teacher Validation Floor & $\tau_{\text{dice}}$ & $0.50$ (DSC gatekeeper) \\
Pseudo-Label Binarization Threshold & $\tau_{\text{bin}}$ & $0.50$ \\
Teacher Focal Loss Weight & $\lambda_{\text{focal}}$ & $0.50$ \\
Temporal Residual Scaling Factor & $\alpha$ & $0.10$ \\
Compressed Bottleneck Width & $C_{\text{temp}}$ & $128$ channels \\
Temporal Smoothness Regularizer & $\beta$ & $0.10$ \\
Sliding Window Sequence Length & $T$ & $11$ frames (stride = 1) \\
Temporal Anchor Frame Loss Weight & $w_{\text{center}}$ & $5.0$ ($w_t = 1.0$ for non-central) \\
Low-Confidence Audit Cutoff & $S_{\text{conf\_thresh}}$ & $0.85$ \\ \hline
\end{tabular}%
}
\end{table}

\subsection{Evaluation Metrics}
Spatial accuracy is evaluated via DSC, NSD@1mm, HD95, and ASSD. Sequence-wide topology is audited via the Anatomical Feasibility Rate:
\begin{equation}
\text{Frac}_{2\text{CC}} = \frac{1}{M} \sum_{m=1}^{M} \mathbb{I}( \text{CCA}(\hat{Y}_m) == 2 )
\end{equation}
which measures the fraction of frames maintaining exactly two foreground components.

\subsection{Quantitative Results and Ablation Study}
We evaluate our spatiotemporal student framework against the fully supervised static 2D teacher, an architectural static student ablation, and a state-of-the-art volumetric baseline framework (nnU-Net). All models are evaluated across 5 folds on the independent, hand-audited validation cohort $\mathcal{D}_T$. Multi-metric quantitative comparisons are detailed in Table~\ref{tab:quantitative_results}.

\begin{table}[t!]
\centering
\caption{Quantitative evaluation and distillation ablation summary across the independent 25-patient test set (948 total frames). Values represent the mean $\pm$ standard deviation across the 5 cross-validation folds. Our model enforces topology.}
\label{tab:quantitative_results}
\resizebox{\textwidth}{!}{%
\begin{tabular}{lccccc}
\hline
\textbf{Architectural Configuration} & $\mathbf{\text{Frac}_{2\text{CC}}}$ \textbf{(\%)} $\uparrow$ & \textbf{NSD@1mm (\%)} $\uparrow$ & \textbf{HD95 (mm)} $\downarrow$ & \textbf{ASSD (mm)} $\downarrow$ & \textbf{DSC (\%)} $\uparrow$ \\ \hline
Static 2D Teacher (Fully-Supervised Base) & $93.0 \pm 4.4$ & $86.6 \pm 5.7$ & $2.88 \pm 0.78$ & $0.58 \pm 0.06$ & $88.8 \pm 2.4$ \\
Static Student & $98.7 \pm 0.7$ & $90.7 \pm 0.4$ & $2.18 \pm 0.09$ & $0.45 \pm 0.03$ & $90.9 \pm 0.3$ \\
\textbf{Spatiotemporal Student (Ours)} & $\mathbf{99.2 \pm 0.6}$ & $\mathbf{92.3 \pm 0.2}$ & $2.05 \pm 0.07$ & $0.41 \pm 0.02$ & $90.9 \pm 0.2$ \\ \hline
2D nnU-Net Baseline & $97.0 \pm 1.6$ & $89.9 \pm 1.4$ & $\mathbf{1.23 \pm 0.36}$ & $\mathbf{0.18 \pm 0.08}$ & $\mathbf{92.9 \pm 1.1}$ \\ \hline
\end{tabular}%
}
\end{table}

\subsubsection{The Inadequacy of Dice Score}

\begin{figure}[t]
\centering
\includegraphics[width=1\textwidth]{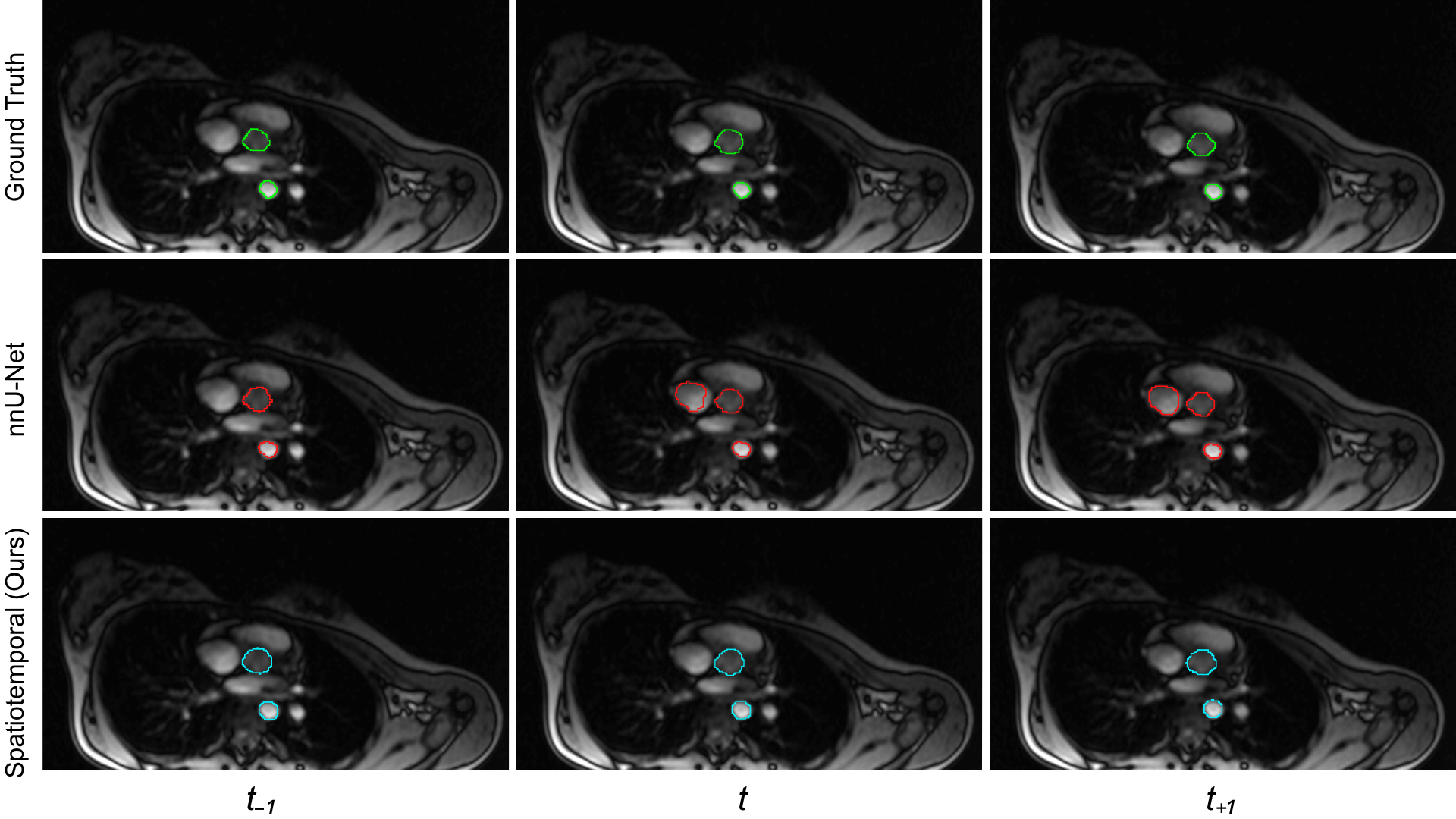}
\caption{Qualitative tracking comparison across consecutive frames ($t-1$, $t$, $t+1$). 2D nnU-Net baseline mistakes a nearby bright vessel for the aorta during a phase of fading visibility ($N=3$). Our spatiotemporal student leverages sequence context to look past this fading effect, maintaining correct anatomical structure ($N=2$).}
\label{fig:qualitative_comparison}
\end{figure}

Although DSC is a standard metric, its bias toward large foreground areas often masks localized boundary errors. As shown in Fig.~\ref{fig:qualitative_comparison}, when aortic visibility fades (frames $t, t+1$), the 2D baseline (Row 2) segments an erroneous structure ($N=3$). Despite this failure, its DSC remains deceptively high due to the high overlap with the ground truth. Conversely, our student model (Row 3) leverages sequence context to look past the fading visibility, maintaining anatomical plausibility ($N=2$) despite the decrease in visibility of the aorta.

\paragraph{Boundary Metrics vs. Topological Validity.}
While 2D nnU-Net achieves lower HD95 ($1.23$~mm vs. $2.05$~mm) and ASSD ($0.18$~mm vs. $0.41$~mm), it does so by optimizing each slice independently, leaving it vulnerable to severe topological failures ($N=3$). In contrast, our student model enforces temporal smoothness ($\beta \mathcal{L}_{\text{smooth}}$) across adjacent frames. This sequence-wide constraint introduces a minor trade-off in local boundary metrics—an increase of $0.23$~mm in ASSD and $0.82$~mm in HD95. Crucially, both deviations remain within approximately one native pixel width ($2.0$~mm resolution). This trade-off is well-justified: preserving $N=2$ topology across the full sequence is essential, as an erroneous extra component completely corrupts downstream cross-sectional area and distensibility calculations.

% =========================================================================
% QUALITATIVE TRACKING COMPARISON GRID
% =========================================================================
% =========================================================================
% =========================================================================

\subsubsection{Large-Scale Inference and Consistency Evaluation}
We used the five cross-validation models to run inference across the unannotated dataset $\mathcal{D}_D$, evaluating $17,539$ continuous frames per fold. An automated audit quantified structural anomalies ($N \neq 2$) and low-confidence frames within the generated segmentations, as detailed in Table~\ref{tab:population_results}.

\begin{table}[t]
\centering
\caption{Large-scale inference consistency and structural anomaly summary across all 5 cross-validation folds. Each entry evaluates fold-specific model performance over the entire unannotated cohort ($17,539$ total frames per run).}
\label{tab:population_results}
\resizebox{\textwidth}{!}{%
\begin{tabular}{lcccccc}
\hline
\textbf{Metric / Configuration} & \textbf{Fold 1} & \textbf{Fold 2} & \textbf{Fold 3} & \textbf{Fold 4} & \textbf{Fold 5} & \textbf{Mean Summary} \\ \hline
\textit{Total Evaluated Frames} & 17,539 & 17,539 & 17,539 & 17,539 & 17,539 & 17,539 Total Frames \\ \hline
\textbf{Static 2D A-UNet Teacher \cite{oktay2018attention}} & & & & & & \\
-- Structural Anomalies ($N \neq 2$) & 1,333 & 2,843 & 1,243 & 1,626 & 1,362 & $1,681.4 \pm 663.0$ \\
-- Low-Confidence Frames & 0 & 837 & 0 & 0 & 0 & $167.4 \pm 374.3$ \\ \hline
\textbf{2D nnU-Net Baseline} & & & & & & \\
-- Structural Anomalies ($N \neq 2$) & 1,169 & 1,115 & 1,089 & 1,109 & 702 & $1,036.8 \pm 190.2$ \\
-- Low-Confidence Frames & 0 & 0 & 0 & 0 & 0 & $\mathbf{0.0 \pm 0.0}$ \\ \hline
\textbf{Static 2D A-UNet Student \cite{oktay2018attention}} & & & & & & \\
-- Structural Anomalies ($N \neq 2$) & 504 & 460 & 456 & 468 & 573 & $492.2 \pm 49.9$ \\
-- Low-Confidence Frames & 8 & 17 & 1 & 22 & 18 & $13.2 \pm 8.6$ \\ \hline
\textbf{Spatiotemporal Student (Ours)} & & & & & & \\
-- Structural Anomalies ($N \neq 2$) & 560 & 383 & 464 & 437 & 412 & $\mathbf{451.2 \pm 67.8}$ \\
-- Low-Confidence Frames & 0 & 0 & 0 & 0 & 0 & $\mathbf{0.0 \pm 0.0}$ \\ \hline
\hline
\end{tabular}%
}
\end{table}

Population-wide evaluation exposes performance gaps hidden by DSC. The Static 2D Teacher exhibits extreme instability, yielding $1,681.4$ structural anomalies per fold. While the nnU-Net baseline eliminates low-confidence predictions, it lacks continuous temporal recurrence, resulting in $1,036.8$ topological anomalies per fold. Our spatiotemporal student achieves the highest stability, suppressing structural violations to $451.2 \pm 67.8$ anomalies per fold and completely eliminating low-confidence tracking failures.

\paragraph{Note on Topological Metrics.} 
Although Phase I pseudo-labels are explicitly cleaned via the offline topological operator, we emphasize that the student network receives no post-processing or topological filtering during test-time inference. Consequently, the high sequence stability ($\text{Frac}_{2\text{CC}} = 99.2\%$) and low population anomaly rate ($451.2$ frames/fold) achieved by the student demonstrate that the model successfully internalizes the $N=2$ structural invariant into its latent spatiotemporal representations, rather than relying on rule-based post-hoc corrections.

\subsubsection{Sensitivity Analysis of Temporal Smoothness Regularization}
We conduct a hyperparameter sensitivity analysis on the temporal smoothness weight $\beta$ (Eq.~\ref{eq:student_loss}) using our population-wide anomaly auditing framework ($17,539$ frames). This sweep balances frame-level voxel overlap against sequence-wide boundary acceleration constraints. Cross-fold mean results across the unannotated cohort are compiled in Table~\ref{tab:lambda_sensitivity}.

\begin{table}[t]
\centering
\caption{Population-wide hyperparameter sensitivity analysis varying the temporal smoothness weight $\beta$ across all $17,539$ unannotated frames.}
\label{tab:lambda_sensitivity}
\resizebox{\textwidth}{!}{%
\begin{tabular}{lccc}
\hline
$\beta$ & \textbf{Anomalies } ($N \neq 2$) $\downarrow$ & \textbf{Low-Conf. Frames } $\downarrow$ & $\text{Frac}_{2\text{CC}}$ \textbf{ (\%)} $\uparrow$ \\ \hline
$0.00$ (Static Base) & $492.2 \pm 49.9$ & $13.2 \pm 8.6$ & $98.7\% \pm 0.7\%$ \\
$0.05$ & $461.2 \pm 60.1$ & $0.0 \pm 0.0$ & $98.9\% \pm 0.6\%$ \\
$0.08$ & $451.8 \pm 64.2$ & $0.0 \pm 0.0$ & $99.0\% \pm 0.7\%$ \\
$\mathbf{0.10}$ & $\mathbf{451.2 \pm 67.8}$ & $\mathbf{0.0 \pm 0.0}$ & $\mathbf{99.2\% \pm 0.6\%}$ \\
$0.15$ & $476.0 \pm 59.2$ & $0.0 \pm 0.0$ & $99.1\% \pm 0.8\%$ \\
$0.50$ & $478.6 \pm 14.7$ & $0.0 \pm 0.0$ & $98.6\% \pm 0.4\%$ \\ \hline
\end{tabular}%
}
\end{table}

Population audits reveal a clear performance peak at $\beta = 0.10$ (Table~\ref{tab:lambda_sensitivity}). Disabling the penalty entirely ($\beta = 0.00$) ignores sequence context and inflates tracking errors. Conversely, over-regularizing the temporal stream ($\beta = 0.50$) induces tracking lag by dampening true biological motion. A balanced penalty is therefore required to filter transient dropouts without restricting natural cardiovascular dynamics.

\section{Conclusion}
This paper introduced a semi-supervised distillation framework for 2D aortic cine-MRI tracking which enforces sequence-wide consistency without requiring extra annotations. Large-scale validation over 17,539 frames demonstrates that our model completely eliminates frame-to-frame tracking dropouts while maintaining high surface accuracy ($\text{NSD@1mm} = 92.3\% \pm 0.2\%$).

A current limitation is that our geometric mask correction strictly assumes two isolated circular cross-sections ($N=2$). While effective for standard short-axis views, this assumption does not hold for complex branching regions (e.g., near the aortic arch) or structural pathologies where the vessel splits. Additionally, the bidirectional ConvLSTM requires buffering an 11-frame window ($T=11$), making the framework suitable for retrospective review rather than real-time streaming. Future work will adapt the method to handle variable topology and online causal tracking. Overall, this framework provides a practical strategy to enforce temporal consistency across dynamic medical imaging tasks.

\section*{Acknowledgement}
The authors would like to express their sincere gratitude to Amr Elsawy and Mohamed Nagy from the Biomedical \& Engineering Innovation Laboratory at the Aswan Heart Centre, Egypt, for their invaluable contributions in providing expert clinical annotations and domain expertise.

\section*{Disclosure of Interests}
The authors have no competing interests to declare that are relevant to the content of this article.

\bibliographystyle{splncs04}
\bibliography{references}

\begin{thebibliography}{10}
\providecommand{\url}[1]{\texttt{#1}}
\providecommand{\urlprefix}{URL }
\providecommand{\doi}[1]{https://doi.org/#1}

\bibitem{bai2018recurrent}
Bai, W., Suzuki, H., Qin, C., Tarroni, G., Oktay, O., Matthews, P.M., Rueckert, D.: Recurrent neural networks for aortic image sequence segmentation with sparse annotations. In: International Conference on Medical Image Computing and Computer-Assisted Intervention (MICCAI). pp. 586--594. Springer (2018)

\bibitem{bratt2021deep}
Bratt, A., Blezek, D.J., Ryan, W.J., Philbrick, K.A., Rajiah, P., Tandon, Y.K., et~al.: Deep learning improves the temporal reproducibility of aortic measurement. Journal of Digital Imaging  \textbf{34}(5),  1183--1189 (2021)

\bibitem{cardoso2022monai}
Cardoso, M.J., Li, W., Brown, R., Ma, N., Kerfoot, E., Wang, Y., Murrey, B., Myronenko, A., Zhao, C., Yang, D., Nath, V., He, Y., Xu, Z., Hatamizadeh, A., Zhu, W., Liu, Y., Zheng, M., Tang, Y., Yang, I., Zephyr, M., Hashemian, B., Alle, S., Darestani, M.Z., Budd, C.: Monai: An open-source framework for deep learning in healthcare. arXiv preprint arXiv:2211.02701  (2022)

\bibitem{cecelja2022aortic}
Cecelja, M., Ruijsink, B., Puyol-Ant{\'o}n, E., Li, Y., Godwin, H., King, A.P., Razavi, R., Chowienczyk, P.: Aortic distensibility measured by automated analysis of magnetic resonance imaging predicts adverse cardiovascular events in {UK Biobank}. Journal of the American Heart Association  \textbf{11}(23),  e026361 (2022)

\bibitem{chen2020deep}
Chen, C., Qin, C., Qiu, H., Tarroni, G., Duan, J., Bai, W., Rueckert, D.: Deep learning for cardiac image segmentation: a review. Frontiers in Cardiovascular Medicine  \textbf{7}, ~25 (2020)

\bibitem{jaffre2023deep}
Jaffr{\'e}, C.M.E.: Deep-{L}earning-based segmentation of the aorta from dynamic {2D} {M}agnetic {R}esonance {I}mages. Master's thesis, Politecnico di Milano (2023)

\bibitem{mosinska2018beyond}
Mosinska, A., M{\'a}rquez-Neila, P., Fua, P.: Beyond the pixel-wise loss for topology-aware delineation. In: IEEE/CVF Conference on Computer Vision and Pattern Recognition (CVPR). pp. 3131--3140 (2018)

\bibitem{oktay2018attention}
Oktay, O., Schlemper, J., Folgoc, L.L., Lee, M., Heinrich, M., Schnabel, J.A., Glocker, B., Rueckert, D.: Attention u-net: Learning where to look for the pancreas. In: International Conference on Medical Imaging with Deep Learning (MIDL) (2018)

\bibitem{painchaud2020cardiac}
Painchaud, N., Skandarani, Y., Judge, T., Bernard, O., Lalande, A., Jodoin, P.M.: Cardiac segmentation with strong anatomical guarantees. IEEE Transactions on Medical Imaging  \textbf{39}(11),  3703--3713 (2020)

\bibitem{paszke2019pytorch}
Paszke, A., Gross, S., Massa, F., Lerer, A., Bradbury, J., Chanan, G., Killeen, T., Lin, Z., Gimelshein, N., Antiga, L., et~al.: Pytorch: An imperative style, high-performance deep learning library. In: Advances in Neural Information Processing Systems (NeurIPS). pp. 8024--8035 (2019)

\bibitem{radosavovic2018data}
Radosavovic, I., Doll{\'a}r, P., Girshick, R., Gkioxari, G., He, K.: Data distillation: Towards omni-supervised learning. In: IEEE/CVF Conference on Computer Vision and Pattern Recognition (CVPR). pp. 4119--4128 (2018)

\bibitem{ronneberger2015u}
Ronneberger, O., Fischer, P., Brox, T.: U-net: Convolutional networks for biomedical image segmentation. In: International Conference on Medical Image Computing and Computer-Assisted Intervention (MICCAI). pp. 234--241. Springer International Publishing (2015)

\bibitem{ruijsink2020fully}
Ruijsink, B., Puyol-Ant{\'o}n, E.., Schnabel, J.A., Razavi, R., King, A.P.: Fully automated, quality-controlled cardiac analysis from {CMR}: Validation and large-scale application to characterize cardiac function. JACC: Cardiovascular Imaging  \textbf{13}(3),  684--695 (2020)

\bibitem{storey2004flow}
Storey, P., Li, W., Chen, Q., Edelman, R.R.: Flow artifacts in steady-state free precession cine imaging. Magnetic Resonance in Medicine  \textbf{51}(1),  115--122 (2004)

\bibitem{tarvainen2017mean}
Tarvainen, A., Valpola, H.: Mean teachers are better role models: Weight-averaged consistency targets improve semi-supervised deep learning results. In: Advances in Neural Information Processing Systems (NeurIPS). vol.~30 (2017)

\bibitem{townsend2015recommendations}
Townsend, R.R., Wilkinson, I.B., Schiffrin, E.L., Avolio, A.P., Chirinos, J.A., Cockcroft, J.R., Heffernan, K.S., Lakatta, E.G., McEniery, C.M., Mitchell, G.F., Najjar, S.S., Nichols, W.W., Urbina, E.M., Weber, T.: Recommendations for improving and standardizing vascular research on arterial stiffness: A scientific statement from the {American} {Heart} {Association}. Hypertension  \textbf{66},  698--722 (2015)

\bibitem{wentland2014review}
Wentland, A.L., Grist, T.M., Wieben, O.: Review of {MRI}-based measurements of pulse wave velocity: a biomarker of arterial stiffness. Cardiovascular Diagnosis and Therapy  \textbf{4}(2),  193--206 (2014)

\end{thebibliography}
\end{document}